# Quantum interference and occupation control in high harmonic generation from monolayer WS$_2$

Minjeong Kim[1,2,†], Taeho Kim[1,2,†], Anna Galler[3,4,†], Dasol Kim[1], Alexis Chacon[1,5,6,7], Xiangxin Gong[8], Yuhui Yang[8], Rouli Fang[8], Kenji Watanabe[9], Takashi Taniguchi[10], B. J. Kim[11], Sang Hoon Chae[8], Moon-Ho Jo[1,2,11], Angel Rubio[4*], Ofer Neufeld[12*], Jonghwan Kim[1,2,11*]

[1]Department of Materials Science and Engineering, Pohang University of Science and Technology, Pohang, Republic of Korea.

[2]Center for van der Waals Quantum Solids, Institute for Basic Science (IBS), Pohang, Republic of Korea.

[3]Institute of Theoretical and Computational Physics, Graz University of Technology, 8010 Graz, Austria.

[4]Max Planck Institute for the Structure and Dynamics of Matter, Hamburg, Germany.

[5]Departamento de Física, Área de Física, Universidad de Panamá, Ciudad Universitaria, nal Physics, Graz UniversitPanama 3366, Panama

[6]Sistema Nacional de Investigación, Building 205, Ciudad del Saber, Clayton Panama, Panama

[7]Parque Científico y Tecnológico, Universidad Autónoma de Chiriquí, Ciudad Universitaria, David 04001, Panama

[8]School of Electrical and Electronics Engineering, School of Materials Science and Engineering, Nanyang Technological University, Singapore, Singapore.

[9]Research Center for Electronic and Optical Materials, National Institute for Materials Science, 1-1 Namiki, Tsukuba 305-0044, Japan

[10]Research Center for Materials Nanoarchitectonics, National Institute for Materials

Science, 1-1 Namiki, Tsukuba 305-0044, Japan

[11]Department of Physics, Pohang University of Science and Technology, Pohang, Republic of Korea.

[12]Technion Israel Institute of Technology, Faculty of Chemistry, Haifa 3200003, Israel.

[†] These authors contributed equally to this work

* To whom correspondence should be addressed: angel.rubio@mpsd.mpg.de (A.R.), ofern@technion.ac.il (O.N.), jonghwankim@postech.ac.kr (J.K.)

**Two-dimensional hexagonal materials such as transition metal dichalcogenides exhibit valley degrees of freedom, offering fascinating potential for valley-based quantum computing and optoelectronics. In nonlinear optics, the K and K' valleys provide excitation resonances that can be used for ultrafast control of excitons, Bloch oscillations, and Floquet physics. Under intense laser fields, however, the role of coherent carrier dynamics away from the K/K' valleys is largely unexplored. In this study, we observe quantum interferences in high harmonic generation from monolayer $WS_2$ as laser fields drive electrons from the valleys across the full Brillouin zone. In the perturbative regime, interband resonances at the valleys enhance high harmonic generation through multi-photon excitations. In the strong-field regime, the high harmonic spectrum is sensitively controlled by light-driven quantum interferences between the interband valley resonances and intraband currents originating from electrons occupying various points in the Brillouin zone, also away from K/K' valleys such as Γ and M. Our experimental observations are in strong agreement with quantum simulations, validating their interpretation. This work proposes new routes for harnessing laser-driven quantum interference in two-dimensional hexagonal systems and all-optical techniques to occupy and read-out electronic structures in the full Brillouin zone via strong-field nonlinear optics, advancing quantum technologies.**

Under intense laser fields solids exhibit extreme nonlinear optical responses such as high-harmonic generation (HHG)[1,2]. Recently, HHG has been demonstrated in diverse material systems – superconductors[3], Mott insulators[4,5], topological solids [6–9] – and has garnered substantial interest as a powerful tool for exploring non-equilibrium quantum phenomena in condensed matter, including Bloch oscillations[10,11], charge coherence[12,13], and phonon dynamics[14–16]. The initial step in HHG is the coherent excitation of electrons from the valence to the conduction bands, forming an electron-hole wave packet [10,11,17–19]. Subsequently, two primary mechanisms contribute to HHG: (1) interband transitions, which induce nonlinear optical polarizations via electron-hole recombination, and (2) intraband transitions, which generate laser-driven anharmonic currents. Two-dimensional (2D) hexagonal materials such as transition metal dichalcogenides provide a fascinating platform for investigating strong-field physics in solids. Due to the degenerate band gaps at the K and K' valleys – where nonzero Berry curvature arises – HHG is intimately connected to valley-specific excitations and all-optical readout[20–24]. Interband excitation and recombination can be resonantly enhanced at the band edges by strong Coulomb interactions in atomically thin 2D structures[20,25–27]. Intraband carrier dynamics also generate nonlinear anomalous and regular currents [28,29], enabling the reconstruction of the Berry curvature[28] and energy dispersions[30–32] across the Brillouin zone (BZ). While bulk materials have shown that strong-field-driven quantum interference between interband and intraband excitation pathways can manipulate HHG[11,33,34], the corresponding effects in 2D hexagonal materials remain largely unexplored.

In the perturbative laser-field regime, local excitation near the K and K' valleys dominate the optical response, offering fundamental physical principles for valleytronic applications that utilize coherent valley-selective excitations as effective two-level systems[35]. Although theoretical models often match experimental data under these conditions, they typically focus only on the valleys and ignore the more intricate band structure away from these high-symmetry points [35]. On the other hand, when the laser field is sufficiently strong, excited carriers in the valleys can travel across the entire BZ, including to non-high-symmetry $k$-points with local energy minima or saddle points. While these alternative $k$-points do not benefit from valley-specific selection rules or minimal bandgap resonances, their coherent excitation under strong-field conditions can manipulate electrons for nonlinear optical applications equivalent to those obtained by valleytronics, i.e. forming a 2-level-like quantum system which is optically manipulable. Fully capturing these dynamics requires methods that drive excitation beyond the K/K´ valleys and identify the unique spectral signatures associated with such extended electron−hole wave packets.

In this study, we investigate how HHG evolves in monolayer $WS_2$ as photo-excited carriers localized within valleys expand throughout the BZ. By irradiating $WS_2$ with mid-infrared pulses at 0.28 eV—resonant with a seven-photon transition to the optical gap—we initiate ultrafast carrier dynamics in the edges of the valence and conduction bands. Systematically increasing the laser intensity causes a transition to the non-perturbative regime, marked by nontrivial spectral features: (1) a pronounced kink in the harmonic yield's intensity dependence, and (2) a distinct spectral evolution that exhibits peak splitting and subsequent merging. Through *ab-initio* and model quantum mechanical simulations, we attribute these phenomena to quantum interference arising from carrier motion transitioning from valley-localized states to highly delocalized states across the Brillouin zone, including the $\Gamma$ and M points, as well as to interference of inter- and intra-band pathways. These findings provide a new mechanism to coherently populate and read out diverse electron–hole superpositions, expanding our ability to manipulate and probe the full BZ of 2D hexagonal solids. Our study thus provides insight that can pave new avenues in ultrafast valleytronics, ultrafast quantum information, and related fields.

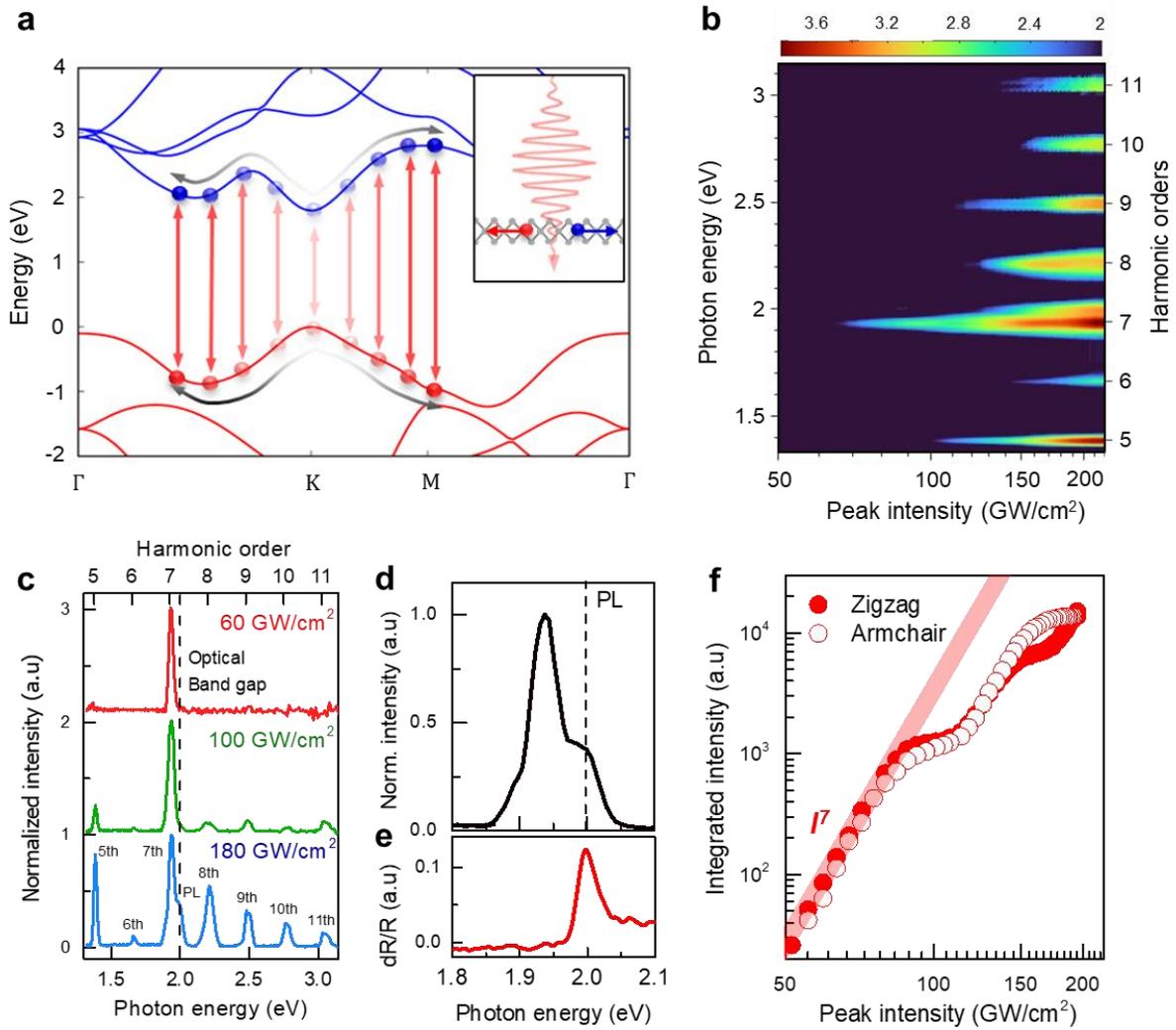

**Fig. 1| Non-perturbative response of lightwave-driven electron-hole pairs in monolayer WS2.** (a) Illustration of the electronic processes under strong laser excitation: Intense laser driving excites carriers throughout the BZ, and drives them in the respective bands. Interband recombination and intraband currents (red and gray arrows, respectively) emit HHG from various points in the BZ, including beyond K/K' valleys. In inset is an illustration of an electrons and hols driven by laser fields. (b) Experimentally measured HHG spectra as a function of laser intensity from 50 to 220 GW/cm². The spectra display markedly different profiles in the perturbative and non-perturbative regimes, showing onset of HHG plateau at higher driving, and the resonant 7$^{th}$ harmonic appearing in much lower intensities. (c) Line-cuts of HHG spectra from (b). (d) Photoluminescence spectra showing clear excitonic signatures with 1s exciton resonance at 2 eV. (e) Reflection contrast spectrum indicating absorption near 2 eV, associated with the 1s exciton resonance. (f) Measured integrated 7$^{th}$ harmonic yield as a function of laser intensity (obtained from (b)), showing a kink feature arising for both zigzag and armchair orientations. Note (f) is plotted in log.

Figure 1a schematically illustrates the electronic processes in monolayer WS$_2$ initiated under intense laser driving. In the perturbative regime nonlinear optical processes primarily arise from excitonic multiphoton transitions at the bandgaps located at the K and K´ valleys. Under intense laser fields, excitons become substantially delocalized through hybridization with higher excitonic bound states and continuum states [36]. Eventually, ionized electrons and holes are driven far beyond the K and K´ valleys over wide regions of momentum space, which opens additional pathways for high-order nonlinear optical processes via intraband and interband transitions.

We fabricate large-area, high-quality, WS$_2$ monolayers using a gold-assisted exfoliation method [37]. To further reduce external defects, the exfoliated monolayer WS$_2$ is encapsulated with hexagonal boron nitride on sapphire substrates via a dry transfer process[38]. Our home-built femtosecond laser system provides linearly-polarized, mid-infrared pulses with ~ 120 fs duration at 100 kHz repetition rate. We intentionally set the photon energy at ~0.28 eV (4500 nm) to match the 7-photon resonant exciton transition at the band edge. The laser power and polarization are precisely controlled and analyzed using polarization optics. HHG spectra are recorded with an electron-multiplying charge-coupled device to achieve a high signal-to-noise ratio. Owing to the combined advantages of high sample quality and high repetition rate, we can sensitively observe the quantum interference features from the perturbative to the non-perturbative regimes, as will be shown below. Further details on the sample preparation and laser set-ups are delegated to the Supplementary Information (SI).

We now explore HHG in monolayer WS$_2$ as a function the driving intensity. HHG spectra exhibit markedly different profiles depending on the laser intensity (see Fig. 1b and 1c). At ~60 GW/cm$^2$ (red solid line in Fig.1c), strong 7$^{th}$ harmonic signals are observed at 1.93 eV, while all other harmonic orders—including lower harmonic orders—are nearly absent. This selective enhancement arises from excitonic resonances at the K and K' valleys. The reflection contrast spectrum (Fig.1e) and HHG spectrum (Fig.1d) show absorption and photoluminescence peaks at 2 eV, respectively, originating from *1s* exciton resonances at the optical gap. The 7$^{th}$ harmonic signal is located near *1s* exciton resonances with a small detuning of 70 meV. Up to ~100 GW/cm$^2$ laser driving, the yield of 7$^{th}$ harmonic in Fig.1f scale as $\propto I^7$ with respect to laser peak intensity (*I*). A similar intensity dependence is observed for photoluminescence under mid-infrared laser excitation (see SI), indicating that all optical processes observed in this regime are primarily mediated by the 7-photon transitions to the resonant excitonic state. Our theoretical calculation provide results in agreement with experiments (which will be discussed below), except for HHG spectra not showing enhancement at the gap without introducing a Coulomb interaction (i.e. in the absence of excitons in the simulation, see further discussion in SI). This further indicates that the resonant enhancement arises from substantial exciton population in the pumped sample.

Figure 1 further shows an HHG plateau spanning 5$^{th}$ to 11$^{th}$ harmonics emerging at higher driving power (> 100 GW/cm$^2$). In these conditions we establish that the HHG yield in non-perturbative, including the 7$^{th}$ harmonic (which is the main observable analyzed in this letter). Notably, Fig.1f shows a pronounced kink at ~120 GW/cm$^2$, which is not expected from the perturbative response, and which we will analyze with theory later on.

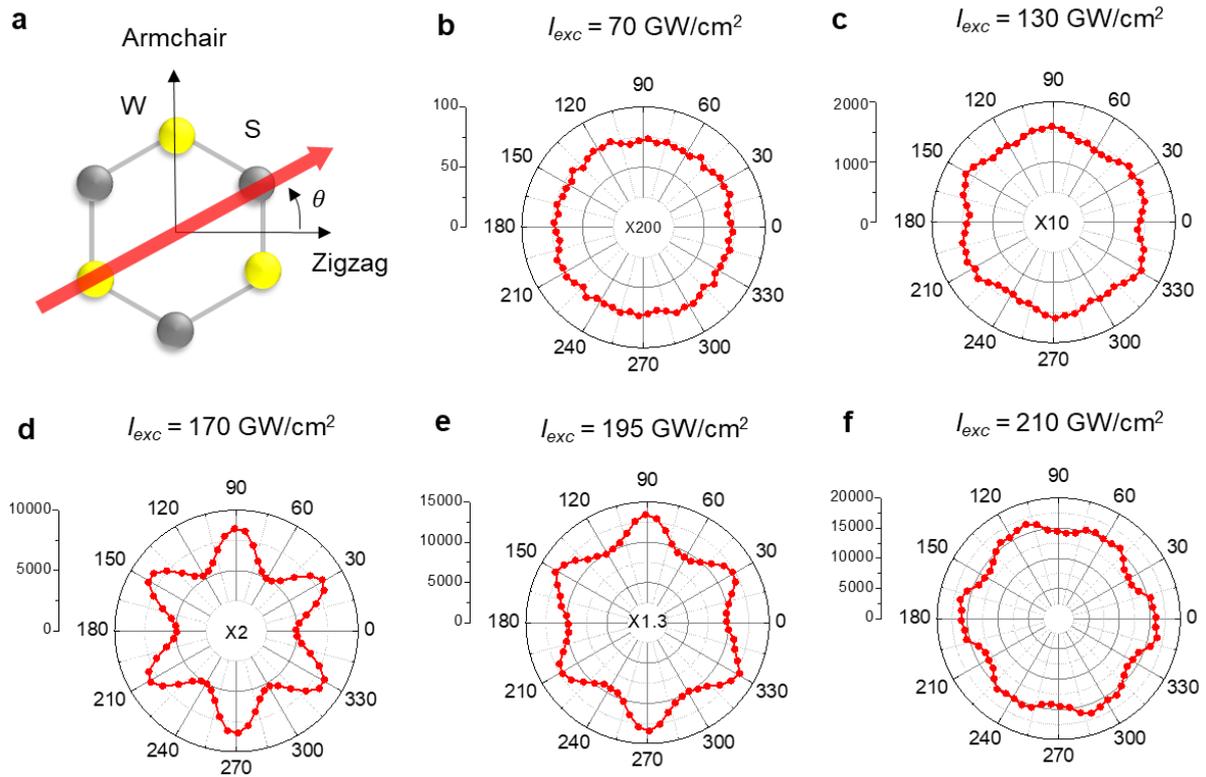

**Fig. 2| Crystal orientation dependence of the 7th harmonic generation yield.** (a) Schematic of the monolayer WS$_2$ crystal structure and laser polarization axis (red arrow). The $x(y)$-axis corresponds to the zigzag (armchair) directions, respectively. The angle θ represents the counterclockwise rotation of the excitation laser polarization relative to the zigzag axis. (b-f) Crystal orientation dependence of seventh harmonics at increasing laser peak intensities: (b) 70GW/cm², (c) 130 GW/cm², (d) 170 GW/cm², (e) 195 GW/cm², and (f) 210GW/cm². For low driving power the harmonic response is isotropic and perturbative. At higher intensity in the transition to non-perturbative HHG, a distinct six-fold pattern emerges with emission along the armchair direction. At yet higher intensity the six-fold pattern is slightly less pronounced and rotates by 30°, exhibiting stronger harmonic intensity along the zigzag direction. The harmonic signals in panels (b), (c), (d), and (e) are magnified by factors of 200, 10, 2, and 1.3, respectively, to clearly visualize the pattern evolution at lower laser intensities.

The yield of the multi-photon resonant 7th harmonic measured as a function of driving orientation can also be indicative of the transition from perturbative to non-perturbative regime. Figures 2b-2f present the integrated yield as a function of the angle between the laser field and the WS$_2$ zigzag direction (see illustration in Fig. 2a). The crystal axis of WS$_2$ is determined ⇄ from polarization analysis on even order harmonics (see SI). Polarization analysis confirms that the 7th harmonic is absent for the polarization component perpendicular to the driving laser field, as expected from HHG dynamical mirror-symmetry selection rules [20,39] (see SI). At a laser intensity of ~70 GW/cm², dominated by perturbative response from excitonic resonances, no apparent dependence on crystal orientation is observed (an isotropic response). However, in the non-perturbative regime harmonic signals exhibit strong 60° periodic modulation, which becomes increasingly pronounced as the laser intensity rises from ~130 to 200 GW/cm², accompanied by significant changes in modulation depth and phase (with 60° periodicity, as expected from crystal symmetry[40]) – Initially, polar plots show stronger harmonic yields along the armchair direction, but as the laser intensity increases the polar plot rotates by 60°, revealing stronger yields along the zigzag direction. Fig.2f is consistent with previous works [20,21] in extreme laser intensities (>1 TW/cm²). The systematic modification of polar plots is indicative of a change in laser excitation regime, and potentially also the physical mechanisms dominating HHG, as will be discussed below.

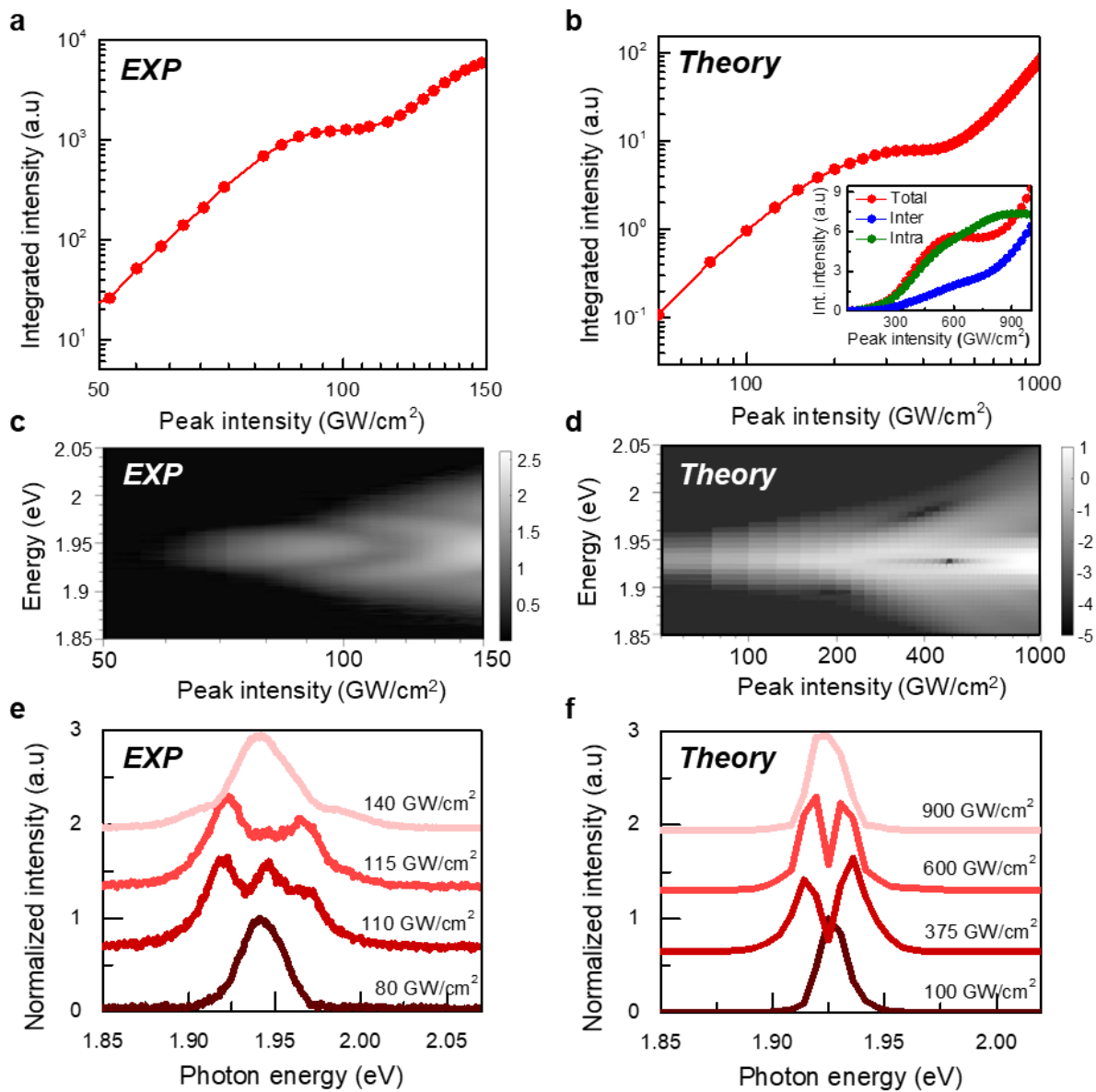

**Fig. 3| Quantum interference in harmonic generation.** (a) 7$^{th}$ harmonic yield vs. driving intensity in zigzag direction (same as Fig. 1(f)). (b) Theoretical calculation of the seventh harmonics as a function of the laser field strength under excitation along zigzag. Inset shows separation to intraband/interband components in linear scale in similar conditions to the experiment demonstrating the kink results from interband/intraband interference. (c) Experimental 2D color map of the seventh harmonic spectra as a function of the laser peak intensity ranging from 50 to 150 GW/cm² under excitation along zigzag orientations. (d) Same as (c) but from quantum mechanical SBE simulations in similar conditions except employing longer pulses for enhanced spectral resolution (see main text). (e) Normalized harmonic spectra at specific laser peak intensity corresponding to 80, 110, 115 and 140 GW/cm², which is linecuts of (c) At low laser field strength, a single peak is observed, but as the field strength increases, this single peak begins to broaden and split into multiple peaks, indicating the emergence of an interference between different electron pathways or transitions. As the field strength further increases to 140 GW/cm² and beyond, the formation of shoulder peaks becomes more pronounced potentially due to more complex quantum pathways or transitions. (f) Theoretical calculation of normalized harmonic spectra at the specific laser peak intensity

corresponding to 100, 375, 600 and 900 GW/cm² , which is the linecut of (d). The calculated spectra exhibit interference patterns that closely resemble the experimental observations, particularly in the peak broadening and the emergence of multiple peaks at higher field strengths. All calculations including were performed using the Tight Binding model, which was employed for solving the Semiconductor Bloch Equations.

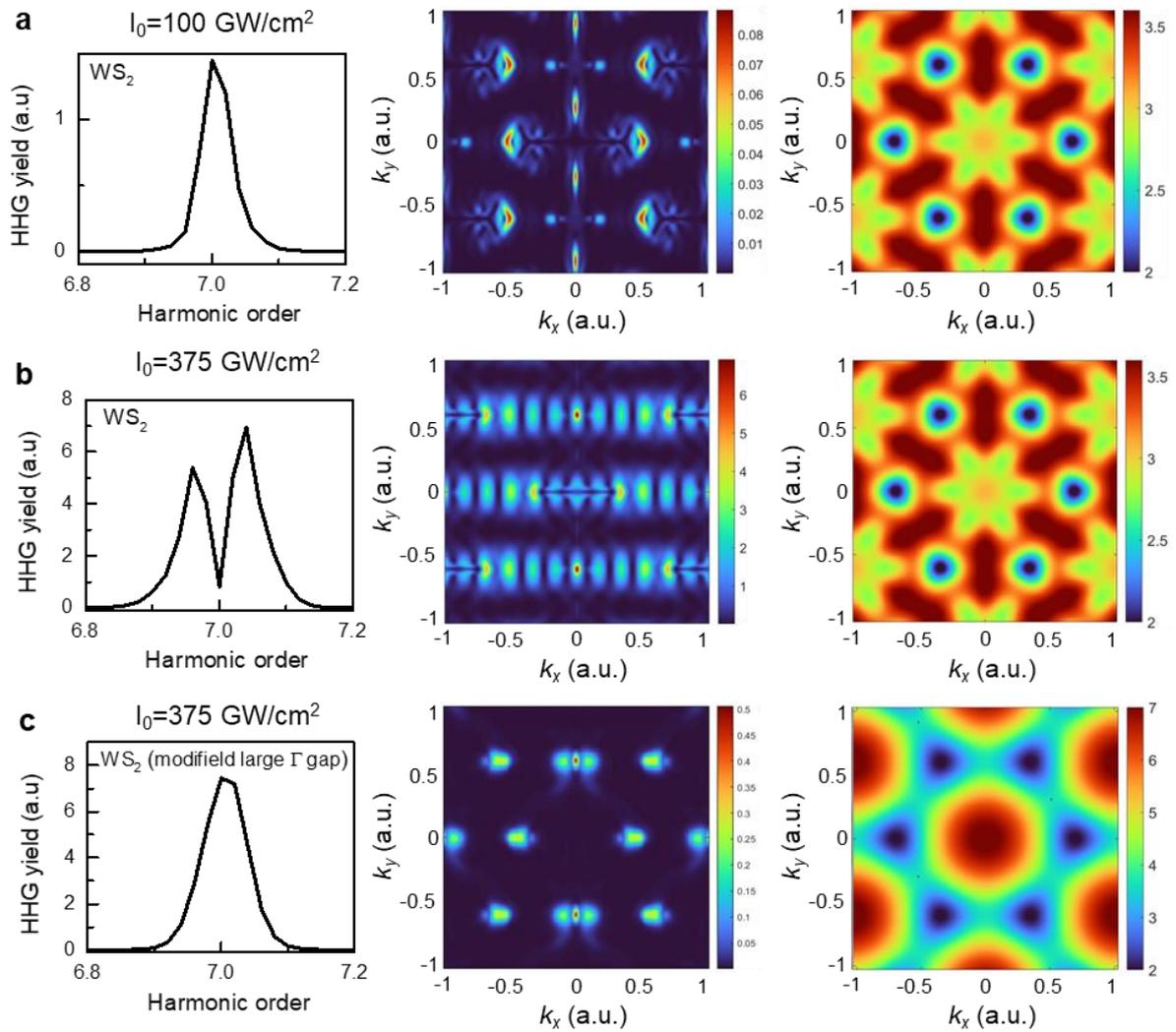

**Fig. 4| Theory of peak splitting in HHG from WS$_2$.** (a) SBE simulated 7$^{th}$ harmonic spectrum (left), showing no onset of peak splitting at lower power (100 GW/cm$^2$). Middle panel shows the k-resolved contributions to this peak, indicating mostly localized charge carrier excitation and emission from the K/K' valleys. Right panel shows the optical gap throughout the BZ in this system. (b) Same as (a) for higher driving power (375 GW/cm$^2$) where there is peak-splitting occurring. Here emission is contributed form delocalized regions in the BZ, including from Γ and M. (c) Same as (b) but with the modified TB model (see text) that reproduced the electronic structure in WS$_2$ only near the K/K' valleys (see right panel), where peak splitting does not occur.

Figure 3a presents a 2D color map of high-resolution recorded 7$^{th}$ harmonic spectra from monolayer WS$_2$ driven in the zig-zag direction as a function of laser intensity (nearly identical spectra are observed along the armchair direction, see SI). There are three key surprising results here, which form the main findings of this letter: (i) At ~100 GW/cm² where non-perturbative responses emerge from the laser-field-driven carriers, a very significant peak broadening arises. (ii) At slightly higher powers (~120 GW/cm$^2$) multiple distinct peaks emerge from the sharp peak that is characteristic of lower intensity driving. At yet higher intensities, ~140 GW/cm², these split peaks converge, resulting in a broader recombined spectral profile. (iii) This evolution is accompanied by a notable kink in the integrated yield of the 7$^{th}$ harmonic (see Fig. 1f and Fig. 3a) over the same laser intensity range, whereby the yield does not increase with increasing driving power.

In the perturbative regime, harmonic spectral profiles are primarily dictated by the driving pulse shape, typically exhibiting Gaussian-like profiles [39]. Beyond the perturbative regime, however, the spectral profile can also be affected by interference between multiple quantum pathways of charge carriers that emerge on sub-laser-cycle timescales. Specifically, destructive interference between distinct quantum pathways can result in peak splitting, reflecting the complex dynamics of charge carriers (as has been observed due to other mechanisms in bulk systems[41–44]). Thus, we hypothesize that these phenomena all arise from multiple quantum path interferences. The main question is then which paths dominate the response of WS$_2$ in this regime?

To address this question, we perform exhaustive theoretical calculations based on several levels of theory. First, *ab-initio* time-dependent density functional theory (TDDFT) simulations are performed and compared with the experiment. Unfortunately, due to the very long-wavelength driving TDDFT fails to reproduce the dominant experimental features. This arises primarily because TDDFT does not include sufficient dephasing channels, which are highly relevant and can significantly alter the HHG spectra in our conditions [45] (because a single driving field period is ~15 fs, meaning dephasing occurs already within a single laser cycle, with recent dephasing times expected to be ~5 fs on average [46]). Nonetheless, the TDDFT simulations allow us to conclude that in our conditions the contribution of electronic correlations and higher/lower-order conduction/valence bands negligibly contribute to the response (see SI). Consequently, we develop a simple two-band model based on a tight-binding (TB) Hamiltonian (with an approach similar to that in refs.[47,48] see SI), which we employ in semiconductor Bloch equations (SBE) in the length gauge in a density matrix formalism[49,50] (given in a.u):

$$\frac{\partial}{\partial t}\rho_{vv}(\mathbf{k},t) = i\mathbf{E}(t) \cdot [\mathbf{d_{cv}}(\mathbf{k})\rho_{cv}^*(\mathbf{k},t) - \mathbf{d_{cv}^*}\rho_{cv}(\mathbf{k},t)]$$

$$\frac{\partial}{\partial t}\rho_{cv}(\mathbf{k},t) = -i\left[\begin{array}{c}\varepsilon_{CB}(\mathbf{k}(t)) - \varepsilon_{VB}(\mathbf{k}(t)) - \frac{i}{T_2} \\ +\mathbf{E}(t) \cdot \left(\begin{array}{c}(\mathbf{d_{cc}}(\mathbf{k}) - \mathbf{d_{vv}}(\mathbf{k}))\rho_{cv}(\mathbf{k},t) \\ +\mathbf{d_{cv}}(\mathbf{k})(2\rho_{vv}(\mathbf{k},t) - 1)\end{array}\right)\end{array}\right] \quad (1)$$

where $\mathbf{k}(t) = \mathbf{k}_0 + \frac{1}{c}\mathbf{A}(t)$, with $\mathbf{k}_0$ the crystal momentum at $t=0$, and $\mathbf{E}(t)$ the electric field vector (in the dipole approximation), which is connected to the vector potential via: $-\partial_t \mathbf{A}(t) = c\mathbf{E}(t)$, and $c$ is the speed of light. In Eq. (1), $\rho_{ij}$ is the density matrix, $\varepsilon_{CB/VB}$ is the band eigen-energy, $\mathbf{d}_{ij}$ are transition dipole matrix elements, and $T_2$ is the phenomenological dephasing time (taken as 5 fs[46]). From the density matrix we obtain the time-dependent current, $\mathbf{J}(t) = \mathbf{J_{intra}}(t) + \mathbf{J_{inter}}(t)$ (separated to inter- intra-band contributions):

$$\mathbf{J_{intra}}(t) = -\sum_{\mathbf{k}\in BZ}[\rho_{vv}(\mathbf{k},t)\mathbf{p_{vv}}(\mathbf{k}(t)) + \rho_{cc}(\mathbf{k},t)\mathbf{p_{cc}}(\mathbf{k}(t))]$$

$$\mathbf{J_{inter}}(t) = -\sum_{\mathbf{k}\in BZ}2\text{Re}[\rho_{cv}(\mathbf{k},t)\mathbf{p_{vc}}(\mathbf{k}(t))] \quad (2)$$

where $\mathbf{p_{ij}}$ are the momentum matrix elements. All momentum and dipole matrix elements, as well as band energies, are obtained through analytical expressions from the TB Hamiltonian, which is optimally fitted to DFT bands throughout across entire BZ with an accurate 14$^{th}$-order nearest-neighbor Hamiltonian (where spin is neglected and with the gap at K/K' offset to match experimental values, as it is often underestimated in DFT). From $\mathbf{J}(t)$ we compute the HHG spectrum as $I_{HHG}(\Omega) = \left|\int dt f(t) \partial_t \mathbf{J}(t) e^{-i\Omega t}\right|^2$, with $f(t)$ being a super-gaussian window function. For all additional technical details of the propagation and numerical procedures see the SI.

Figure 3 presents numerical results employing the SBE-TB formalism showing strong agreement with the experiment – the simulations correctly predict the kink in the 7$^{th}$ harmonic yield vs. power (Fig. 3c and d), and elucidate that this effect arises as a result of interference of interband and intraband emission mechanisms (it is not reproduced by each individual channel, but rather requires their destructive interference). This is the first observation to our knowledge of such clear interferences in 2D systems. We also note that the onset power of this effect is slightly higher in the theory, which likely arises due to excitonic effects not captured in our simulations (i.e. higher intensities are required to pump sufficient excitation in theory, whereas in the experiment larger excitation is mediated through the excitonic resonance). Notably, the theory also establishes that the key experimental findings are fully microscopic in nature.

At a next stage, our theory reproduces the peak broadening and splitting dynamics vs. driving power (see Fig. 3a-d)). Note that here we employed much longer driving laser pulses in order to obtain sufficient spectral resolution (see details in the SI), but otherwise employed the same conditions as in the experiment. Our theoretical analysis reveals that the splitting and converging dynamics do not arise solely due to interference of interband/intraband channels, as the effect appears in each channel separately. To gain further insight, we perform a comprehensive *k*-resolved analysis of the HHG yield, and uncover that at the onset of the peak splitting, a substantial portion of the BZ is excited (comparing occupations in Figs. 4a, 4b, middle panel). Indeed, at high laser powers electrons occupy not only regions near K/K' valleys, but also towards Γ and M points. The HHG emission from these regions is on par with that from the K/K' valleys, and in certain conditions even higher. Mathematically, this is clear due to the relatively low optical gap throughout the BZ (e.g. the gap at Γ is ~3 eV, only ~1 eV higher than the gap at K/K'), but it is also counter-intuitive since most HHG studies in 2D systems only analyze effects near K/K' valleys, which are usually dominant. Overall, this result suggests that the peak splitting arises due to interference of emission from multiple *k*-points.

We validate this conclusion by performing additional simulations where the TB Hamiltonian is modified to reproduce the correct electronic structure only near K/K' valleys, while the gap is artificially increased towards Γ and M to suppress their contribution (see right panels in Fig. 4(b,c)). Indeed, in these conditions the peak splitting phenomena is completely suppressed at identical laser power, corroborating that interference of emission between different points in the BZ accounts for the physical mechanism of peak splitting (and that specifically the K/K' valleys including their Berry curvature cannot alone account for the effect). The peak closing dynamics at yet higher driving is seen to arise due to increased dominance of the intraband emission channel where the split peak converges.

In conclusion, we studied HHG in $WS_2$ monolayers with tunable long-wavelength laser driving. We identified a transition from perturbative HHG—dominated by bound excitons and valley-confined carriers with near-isotropic orientation dependence—to a strong-field regime characterized by delocalized carrier dynamics across the BZ and pronounced anisotropic orientation dependence. In the perturbative regime, exciton resonances strongly enhance the 7$^{th}$ harmonic near the *1s* exciton resonance. As the laser intensity grows, ionized carriers take over, producing nontrivial spectral features such as splitting and multiple kinks in the HHG yield. These experimentally observed, and theoretically validated, phenomena, signify the activation of new quantum pathways in intense fields. Our quantum simulations, including *k*-resolved analysis, reveal that these effects result from quantum interference between interband and intraband transitions, as well as interference from emission between multiple points in the BZ along the laser driving axis. These findings expand our understanding of ultrafast carrier dynamics in valley-based 2D materials and demonstrate the power of HHG for probing light–matter interactions in 2D hexagonal systems.

Especially, we note that these are the first signatures of such interference phenomena in 2D systems, offering a direct all-optical pathway to not only to selectively excite electrons in various high/low-symmetry points of the hexagonal BZ (beyond K/K', also $\Gamma$ and M), but also read them out as clear spectral interference signatures in HHG. Thus, this work paves the way for the next generation of optoelectronic and quantum devices capable of operating at petahertz frequencies, and utilizing multiple *k*-points beyond valleytronics for mimicking 2-level quantum systems.

Methods

**Sample fabrication**

A high-quality monolayer WS$_2$ sample was prepared by the gold-assisted exfoliation method. A 150 nm thickness gold layer was deposited on a flat silicon substrate with a 90 nm thick oxide layer. A polyvinylpyrrolidone (PVP) solution (Sigma Aldrich, mw 40000, 10% wt in ethanol/acetonitrile wt 1/1) was spin-coated on the top of the Au film and cured at 150 °C for 5 minutes. This PVP layer served as a sacrificial layer to prevent contamination from tape residue. The prepared PVP/Au was picked up with thermal release tape (TRT), revealing an ultra-flat, clean, and fresh gold surface-referred to as the gold tape. The gold tape is pressed onto a freshly cleaved bulk WS$_2$ crystal (HQ graphene). As the tape is lifted off the surface, it carries the PVP/Au layer with a monolayer WS2 crystal attached to the Au surface. And then it is further transferred onto a silicon substrate with a 90 nm thick oxide layer. The thermal release tape is removed by heating at 135 °C. The PVP layer is removed by dissolving in deionized (DI) water for 2 hours. Finally, the sample on the substrate, covered by Au layer, was rinsed with acetone and cleaned by O2 plasma for 4 minutes to remove any remaining polymer residues.

The van der Waals heterostructure of WS$_2$ monolayer and hBN was prepared by the dry transfer technique [38]. Thickness of WS$_2$ monolayer was first identified by the optical contrast of a microscope image, followed by the detailed spectroscopic characterization. Approximately 20 nm-thick hBN flakes were exfoliated onto a silicon substrate with 90 nm oxide layer. To fabricate the encapsulated WS$_2$ monolayer, we used the thermoplastic methacrylate copolymer (Elvacite 2552C, Lucite International) stamp to pick up the hBN flakes and WS$_2$ monolayer in sequence with accurate alignment based on an optical microscope. The Elvacite stamp with the heterostructure was then stamped onto a sapphire substrate. The polymer and samples were heated up at 70 °C for the pick-up and 200 °C for the stamp process, respectively. Finally, we dissolved the Elavcite in acetone for 3 minutes at 100 °C.

**HHG measurements**

Mid-infrared pulses were generated from a femtosecond laser system (Light Conversion PHAROS) using an optical parametric amplifier (ORPHEUS) and a difference frequency generator (LYRA). The output served wavelength-tunable multi-cycle pulses with a repetition rate of 100 kHz. The spectral linewidth of the pulse was 15.4 meV in full width at half maximum (FWHM), and the pulse duration was estimated to be 120 fs, assuming a Fourier-transform-limited pulse. Laser intensity was precisely controlled by a pair of linear polarizers inserted into the beam path. Half-wave plates were also inserted into the beam path to control the polarization of the excitation laser. The mid-infrared pulses were then focused near the center of the monolayer $WS_2$ sample using ZnSe focusing objectives, producing a spot size of approximately 60–80 µm. The emitted HHG was collected by a 50× objective lens in a transmission geometry, and its polarization was analyzed by a half-wave plate mounted on a motorized stage and a fixed Glan-Taylor polarizer. The HHG spectra were recorded by an electron-multiplying charge-coupled device detector (ProEM, Princeton Instruments) and a grating spectrometer (SP-2300, Princeton Instruments) at Materials Imaging & Analysis Center of POSTECH.

**References**

1. Ghimire, S. *et al.* Observation of high-order harmonic generation in a bulk crystal. *Nature Phys* **7**, 138–141 (2011).

2. Ghimire, S. & Reis, D. A. High-harmonic generation from solids. *Nature Phys* **15**, 10–16 (2019).

3. Alcalà, J. *et al.* High-harmonic spectroscopy of quantum phase transitions in a high-Tc superconductor. *Proceedings of the National Academy of Sciences* **119**, e2207766119 (2022).

4. Orthodoxou, C., Zaïr, A. & Booth, G. H. High harmonic generation in two-dimensional Mott insulators. *npj Quantum Mater.* **6**, 76 (2021).

5. Murakami, Y., Eckstein, M. & Werner, P. High-Harmonic Generation in Mott Insulators. *Phys. Rev. Lett.* **121**, 057405 (2018).

6. Bai, Y. *et al.* High-harmonic generation from topological surface states. *Nat. Phys.* **17**, 311–315 (2021).

7. Heide, C. *et al.* Probing topological phase transitions using high-harmonic generation. *Nat. Photon.* **16**, 620–624 (2022).

8. Schmid, C. P. *et al.* Tunable non-integer high-harmonic generation in a topological insulator. *Nature* **593**, 385–390 (2021).

9. Neufeld, O., Tancogne-Dejean, N., Hübener, H., De Giovannini, U. & Rubio, A. Are There Universal Signatures of Topological Phases in High-Harmonic Generation? Probably Not. *Phys. Rev. X* **13**, 031011 (2023).

10. Reislöhner, J., Kim, D., Babushkin, I. & Pfeiffer, A. N. Onset of Bloch oscillations in the almost-strong-field regime. *Nat Commun* **13**, 7716 (2022).

11. Schubert, O. *et al.* Sub-cycle control of terahertz high-harmonic generation by dynamical Bloch oscillations. *Nature Photon* **8**, 119–123 (2014).

12. Heide, C. *et al.* Probing electron-hole coherence in strongly driven 2D materials using high-harmonic generation. *Optica* **9**, 512–516 (2022).

13. Freudenstein, J. *et al.* Attosecond clocking of correlations between Bloch electrons. *Nature* **610**, 290–295 (2022).

14. Neufeld, O., Zhang, J., De Giovannini, U., Hübener, H. & Rubio, A. Probing phonon dynamics with multidimensional high harmonic carrier-envelope-phase spectroscopy. *Proceedings of the National Academy of Sciences* **119**, e2204219119 (2022).

15. Zhang, J. *et al.* High-harmonic spectroscopy probes lattice dynamics. *Nat. Photon.* **18**, 792–798 (2024).

16. Rana, N., Mrudul, M. S., Kartashov, D., Ivanov, M. & Dixit, G. High-harmonic spectroscopy of coherent lattice dynamics in graphene. *Phys. Rev. B* **106**, 064303 (2022).

17. Wu, M., Ghimire, S., Reis, D. A., Schafer, K. J. & Gaarde, M. B. High-harmonic generation from Bloch electrons in solids. *Phys. Rev. A* **91**, 043839 (2015).

18. Vampa, G. *et al.* Theoretical Analysis of High-Harmonic Generation in Solids. *Phys. Rev. Lett.* **113**, 073901 (2014).

19. Yue, L. & Gaarde, M. B. Introduction to theory of high-harmonic generation in solids: tutorial. *J. Opt. Soc. Am. B* **39**, 535–555 (2022).

20. Yoshikawa, N. *et al.* Interband resonant high-harmonic generation by valley polarized electron–hole pairs. *Nat Commun* **10**, 3709 (2019).

21. Liu, H. *et al.* High-harmonic generation from an atomically thin semiconductor. *Nature Phys* **13**, 262–265 (2017).

22. Jiménez-Galán, Á., Silva, R. E. F., Smirnova, O. & Ivanov, M. Sub-cycle valleytronics: control of valley polarization using few-cycle linearly polarized pulses. *Optica* **8**, 277–280 (2021).

23. Mrudul, M. S., Jiménez-Galán, Á., Ivanov, M. & Dixit, G. Light-induced valleytronics in pristine graphene. *Optica* **8**, 422 (2021).

24. Neufeld, O., Hübener, H., Jotzu, G., De Giovannini, U. & Rubio, A. Band Nonlinearity-Enabled Manipulation of Dirac Nodes, Weyl Cones, and Valleytronics with Intense Linearly Polarized Light. *Nano Lett.* **23**, 7568–7575 (2023).

25. Hader, J., Neuhaus, J., Moloney, J. V. & Koch, S. W. Coulomb enhancement of high harmonic generation in monolayer transition metal dichalcogenides. *Opt. Lett.* **48**, 2094 (2023).

26. Molinero, E. B. *et al.* Subcycle dynamics of excitons under strong laser fields. *Science Advances* **10**, eadn6985.

27. Chang Lee, V., Yue, L., Gaarde, M. B., Chan, Y. & Qiu, D. Y. Many-body enhancement of high-harmonic generation in monolayer MoS2. *Nat Commun* **15**, 6228 (2024).

28. Luu, T. T. & Wörner, H. J. Measurement of the Berry curvature of solids using high-harmonic spectroscopy. *Nat Commun* **9**, 916 (2018).

29. Yue, L. & Gaarde, M. B. Characterizing Anomalous High-Harmonic Generation in Solids. *Phys. Rev. Lett.* **130**, 166903 (2023).

30. Vampa, G. *et al.* All-Optical Reconstruction of Crystal Band Structure. *Phys. Rev. Lett.* **115**, 193603 (2015).

31. Borsch, M. *et al.* Super-resolution lightwave tomography of electronic bands in quantum materials. *Science* **370**, 1204–1207 (2020).

32. Lanin, A. A., Stepanov, E. A., Fedotov, A. B. & Zheltikov, A. M. Mapping the electron band structure by intraband high-harmonic generation in solids. *Optica* **4**, 516 (2017).

33. Hohenleutner, M. *et al.* Real-time observation of interfering crystal electrons in high-harmonic generation. *Nature* **523**, 572–575 (2015).

34. Tancogne-Dejean, N., Mücke, O. D., Kärtner, F. X. & Rubio, A. Ellipticity dependence of high-harmonic generation in solids originating from coupled intraband and interband dynamics. *Nat Commun* **8**, 745 (2017).

35. Schaibley, J. R. *et al.* Valleytronics in 2D materials. *Nature Reviews Materials* **1**, 16055 (2016).

36. Kobayashi, Y. *et al.* Floquet engineering of strongly driven excitons in monolayer tungsten disulfide. *Nat. Phys.* (2023) doi:10.1038/s41567-022-01849-9.

37. Liu, F. *et al.* Disassembling 2D van der Waals crystals into macroscopic monolayers and reassembling into artificial lattices. *Science* **367**, 903–906 (2020).

38. Wang, L. *et al.* One-Dimensional Electrical Contact to a Two-Dimensional Material. *Science* **342**, 614–617 (2013).

39. Shen, Y. R. *The Principles of Nonlinear Optics*. (Wiley, 2003).

40. Neufeld, O., Podolsky, D. & Cohen, O. Floquet group theory and its application to selection rules in harmonic generation. *Nature Communications* **10**, 405 (2019).

41. Xia, P. *et al.* High-harmonic generation in GaAs beyond the perturbative regime. *Phys. Rev. B* **104**, L121202 (2021).

42. Sekiguchi, F. *et al.* Enhancing high harmonic generation in GaAs by elliptically polarized light excitation. *Phys. Rev. B* **108**, 205201 (2023).

43. Reislöhner, J., Kim, D., Babushkin, I. & Pfeiffer, A. N. Onset of Bloch oscillations in the almost-strong-field regime. *Nat Commun* **13**, 7716 (2022).

44. Kim, Y. W. *et al.* Spectral Interference in High Harmonic Generation from Solids. *ACS Photonics* **6**, 851–857 (2019).

45. Kim, Y. *et al.* Dephasing Dynamics Accessed by High Harmonic Generation: Determination of Electron–Hole Decoherence of Dirac Fermions. *Nano Lett.* **24**, 1277–1283 (2024).

46. Korolev, V. *et al.* Unveiling the Role of Electron-Phonon Scattering in Dephasing High-Order Harmonics in Solids. *arXiv preprint arXiv:2401.12929* (2024).

47. Wang, Y. *et al.* Tight-binding model for electronic structure of hexagonal boron phosphide monolayer and bilayer. *Journal of Physics: Condensed Matter* **31**, 285501 (2019).

48. Galler, A., Rubio, A. & Neufeld, O. Mapping Light-Dressed Floquet Bands by Highly Nonlinear Optical Excitations and Valley Polarization. *J. Phys. Chem. Lett.* **14**, 11298–11304 (2023).

49. Yue, L. & Gaarde, M. B. Structure gauges and laser gauges for the semiconductor Bloch equations in high-order harmonic generation in solids. *Phys. Rev. A* **101**, 053411 (2020).

50. Mrudul, M. S. & Dixit, G. High-harmonic generation from monolayer and bilayer graphene. *Phys. Rev. B* **103**, 094308 (2021).

Acknowledgement: We acknowledge fruitful discussions with Prof. Dieter Bauer, Prof. Marcelo Ciappinna, Prof. Gopal Dixit, and Dr. Lun Yue. This work was supported by the Institute for Basic Science (IBS), Korea under Project Code IBS-R014-A1. J.K. acknowledge the support from the National Research Foundation of Korea grants (NRF-2023R1A2C2007998). This study was also supported by the MSIT(Ministry of Science and ICT), Korea, under the ITRC (Information Technology Research Center) support program (IITP-2023-RS-2022-00164799) supervised by the IITP(Institute for Information & Communications Technology Planning & Evaluation). This work was supported by the European Research Council (ERC-2015-AdG694097), the Cluster of Excellence 'Advanced Imaging of Matter' (AIM), Grupos Consolidados (IT1453-22), and Deutsche Forschungsgemeinschaft (DFG)-SFB-925-proj-ect 170620586. The Flatiron Institute is a division of the Simons Foundation. We acknowledge support from the Max Planck-New York City Center for Non-Equilibrium Quantum Phenomena. A.C. acknowledges partial support by the Sistema Nacional de Investigación de Panama. A.G. acknowledges support by the Austrian Science Fund (FWF) grant 10.55776/V988. S.H.C. acknowledges funding from the A*STAR, Singapore, Advanced Manufacturing and Engineering (AME) Individual Research Grant (IRG) under the Project M23M6c0109. This work is supported by the MOE AcRF Tier 3 grant (MOE-MOET32023-0003) "Quantum Geometric Advantage" and the Nanyang NanoFabrication Centre (N2FC). K.W. and T.T. acknowledge support from the JSPS KAKENHI (Grant Numbers 21H05233 and 23H02052) , the CREST (JPMJCR24A5), JST and World Premier International Research Center Initiative (WPI), MEXT, Japan.

Competing interests: The authors declare no competing interests.

Supplementary Information for

# Quantum interference and occupation control in high harmonic generation from monolayer $WS_2$

**Contents**

**Supplementary Figures (Figure S1 – S5)**

1. Power dependence of photoluminescence signals under mid-infrared laser excitation
2. Comparison of HHG Spectra with and without Coulomb Interaction
3. Decomposition of seventh harmonic generation and photoluminescence from emission spectrum with polarization analysis
4. Crystal orientation dependence of even order harmonics under mid-infrared laser excitation
5. Quantum interference in harmonic generation under mid-infrared laser excitation along zigzag and armchair

**Supplementary Note**

  Sample fabrication

  Reflection contrast spectroscopy

  HHG measurements (Figure S6)

  TDDFT simulations (Figure S7 - S8)

Supplementary Figures

# Figure S1. Power dependence of photoluminescence signals under mid-infrared laser excitation

Fig. S1 shows the emission spectrum of monolayer $WS_2$ under intense mid-infrared laser excitation, showing both photoluminescence and seventh harmonic generation signals as a function of the laser peak intensity. We set our laser excitation wavelength to 4500nm to clearly distinguish $7^{th}$ harmonic signals and PL signals. Notably, the seventh harmonic exhibits spectral broadening and peak splitting as laser peak intensity increases, while the photoluminescence intensity also shows a consistent increasing trend in Fig. S1(a). By decomposing the spectrum into seventh harmonic and photoluminescence components, we can clearly observe the trends shown in Fig. S1(b) and (c). Details of the decomposition method will be discussed further in Fig. S3. As shown in Fig. S1(b), the seventh harmonic intensity follows a $I^7$ scaling law up to approximately 100 GW/cm², characteristic of the perturbative regime. Beyond this regime, multiple kinks become apparent, which is similar behavior in the main text. The photoluminescence data in Fig. S1(c) initially follows a power law scaling near approximately 100 GW/cm², consistent with carrier generation through multiphoton absorption process in the perturbative regime. However, as the laser intensity increases, the photoluminescence signal deviates from the power-law scaling..

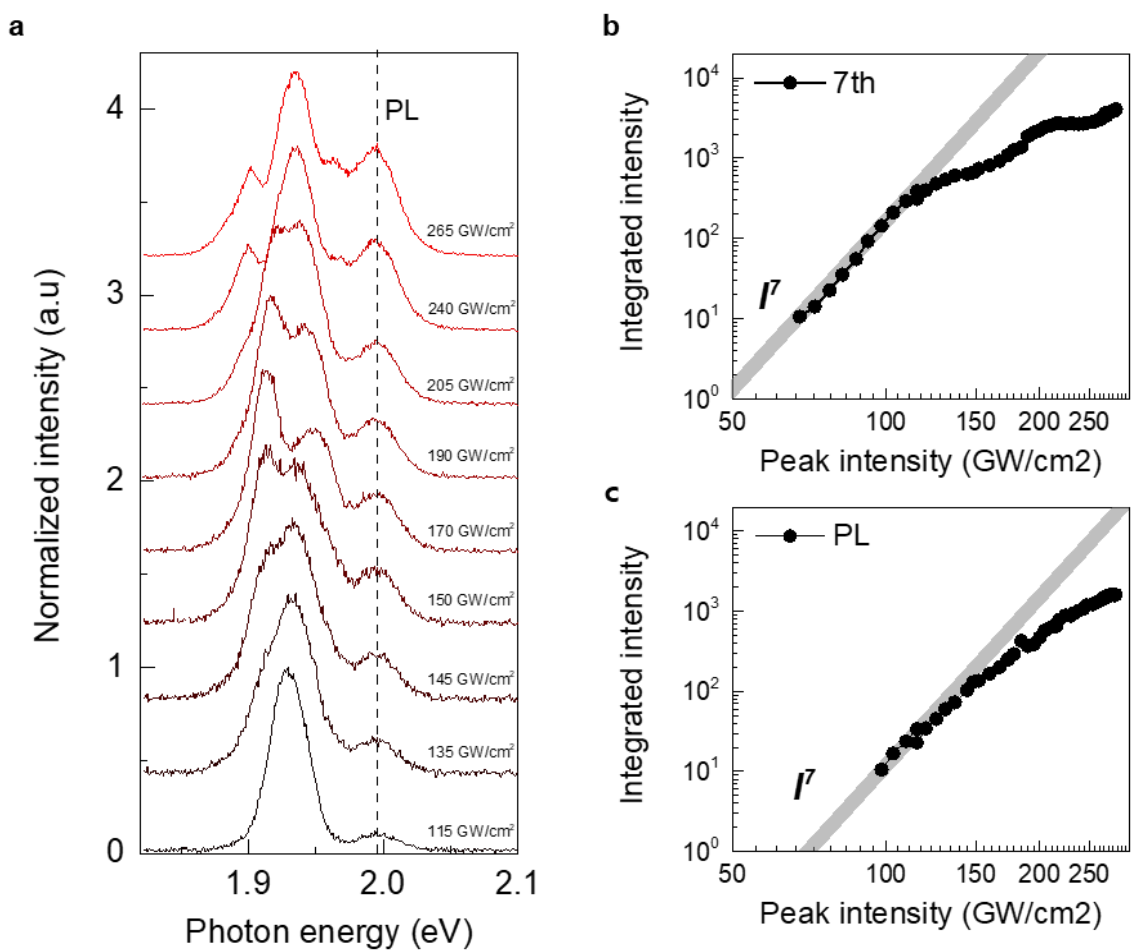

**Figure S1| Power dependence of photoluminescence signals under mid-infrared laser excitation.** (a) Photoluminescence (PL) and the seventh harmonic spectra from monolayer $WS_2$ under mid-infrared laser excitation at varying peak intensities from 115 GW/cm$^2$ to 265 GW/cm$^2$. At 115 GW/cm², the seventh harmonic (1.93 eV) and PL (1.99 eV) are clearly observed, and the spectral broadening and peak splitting of the seventh harmonic are confirmed as the laser peak intensity increases. (b) The seventh harmonic and (c) PL intensity as a function of the laser peak intensity. The seventh harmonic intensity initially scales with the seventh power of the laser peak intensity (gray line) via multiphoton absorption process. The seventh harmonic also exhibits multiple kinks in peak intensity dependence beyond the perturbative regime.

**Figure S2. Comparison of HHG Spectra with and without Coulomb Interaction**

Coulomb interaction plays a pivotal role in the HHG process. As shown in Fig. S2(a), the experimental data exhibits a significant enhancement of the seventh harmonic, surpassing the intensity of the fifth harmonic. This enhancement is driven by the excitonic resonance near the seventh harmonic energy. Furthermore, the absorption at this energy is resonantly enhanced, leading to increased multi photon carrier that further contributes to the HHG yield. In contrast, Fig. S2(b) shows calculated spectra without Coulomb interaction, where the seventh harmonic is significantly weaker than the fifth harmonic. This behavior is consistent with the conventional expectation in HHG, where harmonic yield substantially decreases as the harmonic order increases. The difference between the two spectra highlights the critical role of Coulomb interactions in enhancing HHG at specific orders.

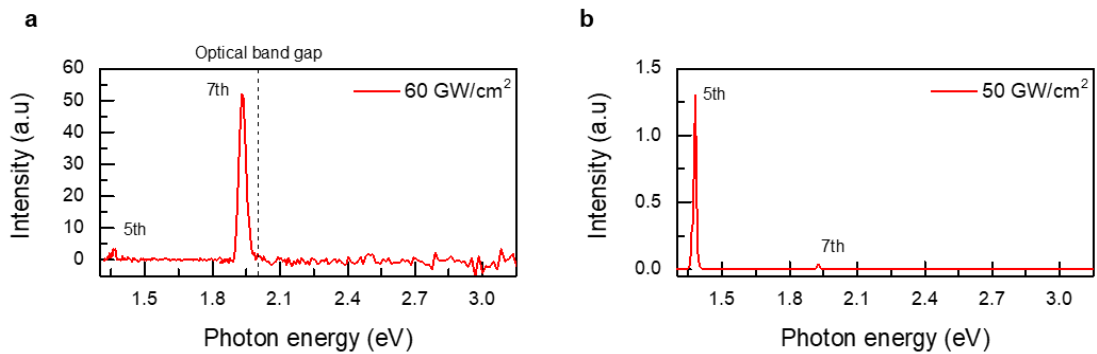

**Figure S2| Experimental and theoretical calculated harmonic spectra.** (a) Experimental harmonic spectra under laser peak intensity 60 GW/cm$^2$. Coulomb interaction plays a significant role in enhancing harmonic intensities near exciton resonance. (b) Theoretical calculation of harmonic spectra under laser peak intensity 50 GW/cm$^2$.

**Figure S3. Decomposition of seventh harmonic generation and photoluminescence from emission spectrum with polarization analysis**

7$^{th}$ harmonic signals in monolayer TMDs can be accurately extracted in the emission spectra by utilizing polarization dependence of photoluminescence (PL) responses. As shown in Fig. S3, when measuring in the parallel-polarized configuration, both the seventh harmonic and PL signals are present. However, in the cross-polarized configuration, only the PL and even-order harmonics appear, while the seventh harmonic is absent. This polarization dependence allows the separation of PL contributions from the seventh harmonic spectra. To obtain the pure seventh harmonic signal presented in the main text, the cross-polarized measurements (the PL contribution) were subtracted from the parallel-polarized measurements. The polarization-resolved emission spectra in Fig. S3(a) confirms that the seventh harmonic at 1.93 eV appears only in the parallel configuration, whereas the eighth harmonic at 2.2 eV appears only in the cross configuration, alongside the PL peak at approximately 2.0 eV. Additionally, to verify that the seventh harmonic is completely absent in the cross-polarized configuration, crystal orientation-dependent measurements were performed in both polarization settings. As shown in Fig. S3(b), the seventh harmonic exhibits a distinct six-fold pattern in the parallel configuration, while in the cross configuration, the signal remains at the noise level. Based on this confirmation, the analysis in the main text focuses exclusively on the parallel-polarized seventh harmonic data.

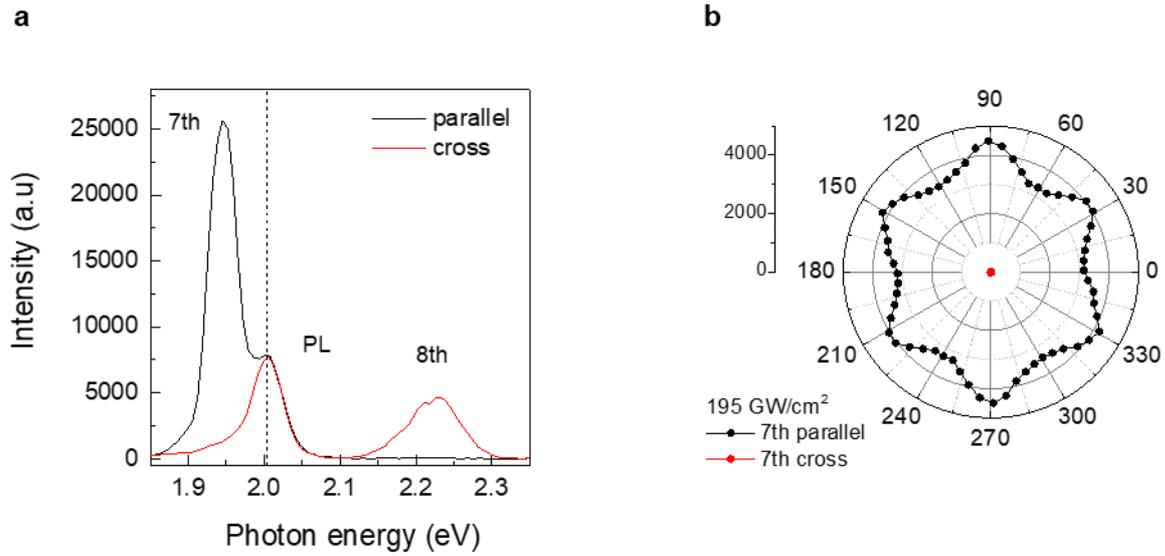

**Figure S3| Polarization of the high harmonic generation from monolayer WS$_2$ under mid-infrared laser excitation.** (a) Emission spectra from monolayer WS$_2$ in both parallel- (black) and cross- (red) polarized configurations. The PL peak at approximately 2.0 eV appears in both parallel and cross-polarized configurations. The odd order harmonics are present only in the parallel configuration (black), with the seventh harmonic at 1.93 eV. In contrast, the even-order harmonics are observed only in the cross-polarized configuration (red), with the eighth harmonic at 2.2 eV. (b) Crystallographic orientation dependence of the seventh harmonics in parallel- (black) and cross- (red) polarized configurations, showing a clear six-fold pattern aligned with zigzag direction at a laser intensity of 195 GW/cm².

**Figure S4. Crystal orientation dependence of even order harmonics under mid-infrared laser excitation**

High-harmonic generation (HHG) in monolayer TMDs provides unique insights into the interplay between crystal symmetry and light-matter interactions. The polarization dependence of HHG, as demonstrated in Fig. S3 arises due to the strong coupling between the crystal symmetry and the fundamental laser. For excitation along the zigzag direction, the mirror symmetry plane normal to this axis fundamentally shapes the electron-hole dynamics. When the laser field reaches its peak, electrons are excited from the valence to the conduction band, primarily at the $K$ and $K'$ valleys in the Brillouin zone. These excited carriers undergo asymmetric acceleration and recombination dynamics due to the broken inversion symmetry and the influence of the crystal's band structure. This asymmetry enforces a temporal behavior in which the interband polarization density $p(t)$ [1] flips its sign every half-cycle of the laser field, satisfying $p(t+T/2) = -p(t)$, where $T$ is the laser field period. As a result of this symmetry constraint, odd-order harmonics are emitted with polarization parallel to the fundamental laser field, while even-order harmonics are polarized perpendicular. Similarly, for excitation along the armchair direction, the polarization of HHG can be analyzed, resulting in both even- and odd-order HHG emission appearing with parallel polarization in the case of armchair excitation. In the weak laser intensity, it was experimentally confirmed that odd-order harmonics appear only in parallel configuration, while even-order harmonics are observed only in the cross configuration.

We performed crystal orientation-dependent measurements of even-order harmonics under cross-polarized detection. As shown in Fig. S4, the eighth and tenth harmonics exhibit clear six-fold symmetric patterns, confirming the selection rules dictated by the monolayer $WS_2$ crystal symmetry. The orientation dependence of the even-order harmonics provides a robust reference for sample alignment, ensuring the consistency of our experimental setup. Each sample's experimental angle was determined using these even-order harmonic measurements, allowing reliable interpretation of the results. The observed six-fold symmetry further highlights the interplay between the crystal's mirror symmetry and the laser field polarization, governing the polarization selection rules for even-order harmonics.

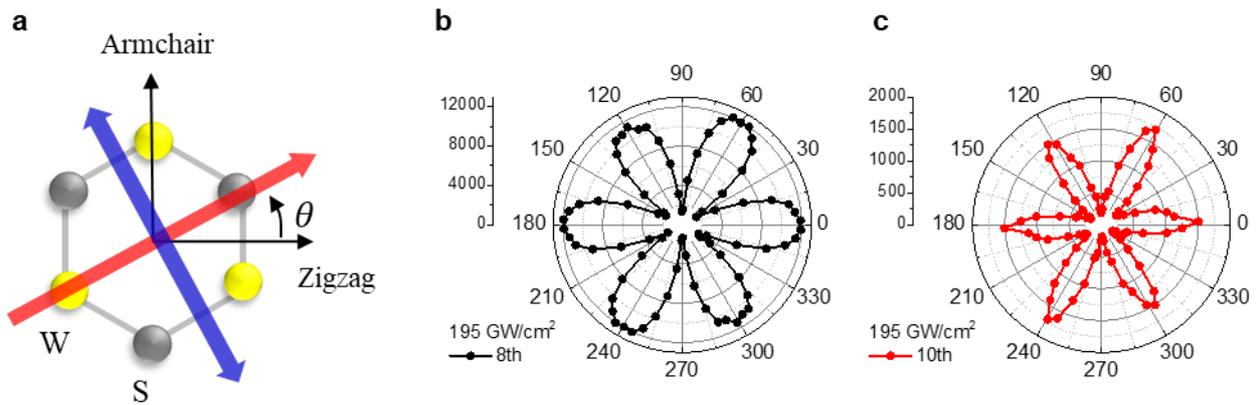

**Figure S4| Crystal orientation dependence of even order harmonics under cross configuration.** (a) Schematic of the monolayer $WS_2$ crystal structure, showing the zigzag and armchair directions. The red arrow indicates the orientation of the excitation laser polarization, rotated by an angle $\theta$ counterclockwise from the zigzag direction, while the blue arrow represents the detection direction in the cross-polarized configuration. Crystal orientation dependence of the eighth (b) and tenth (c) harmonic yields under cross-polarized configuration at a laser intensity of 195 GW/cm². Both harmonics exhibit distinct sixfold symmetric patterns.

**Figure S5. Quantum interference in harmonic generation under mid-infrared laser excitation along zigzag and armchair**

We investigated the seventh harmonic generation under mid-infrared laser excitation along both zigzag and armchair orientation. For excitation along the zigzag direction (Fig. S5 (a)), as described in the main text, the seventh harmonic demonstrates a clear evolution with increasing intensity. At lower intensities, it appears as a single peak, while at higher intensities, the peak broadens and exhibits splitting, indicating the activation of additional quantum pathways. Similarly, in the armchair direction (Fig.S5 (b)), a comparable trend is observed, with spectral broadening and splitting becoming apparent as the intensity increases, suggesting similar underlying physical mechanisms in both orientations. Although the excitation direction influences the electron dynamics in momentum space, resulting in different carrier trajectories. The spectral features suggest that, from the perspective of quantum interference, similar phenomena occur in both orientations. To further analyze these spectral dynamics, selected harmonic spectra corresponding to vertical line cuts of the 2D color maps are shown in Fig. S5 (c,d) for the zigzag and armchair. These spectra clearly demonstrate the transition from a perturbative regime, where harmonics appear as well-defined peaks, to a strong-field regime characterized by spectral broadening and the emergence of additional features. Although the excitation direction influences electron dynamics in momentum space, leading to distinct carrier trajectories. The similarities in spectral evolution suggest that, from the perspective of quantum interference, analogous mechanisms govern the harmonic generation process in both orientations.

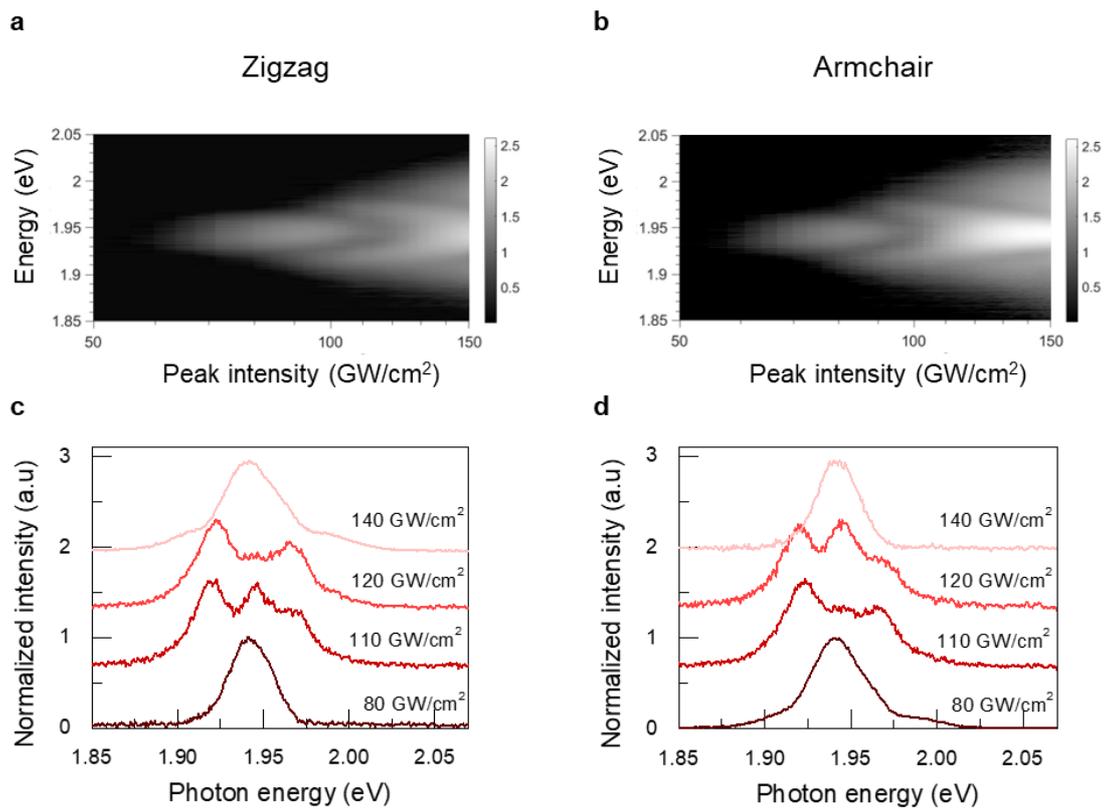

**Figure S5| Harmonic spectra under mid-infrared laser excitation.** Experimental 2D color map of the seventh harmonic spectra from monolayer $WS_2$ as a function of peak intensity ranging from 80 to 140 GW/cm² under excitation along the zigzag (a) and armchair (b) crystallographic orientations. In both orientations, the spectra exhibit distinct signatures of quantum interference as the intensity increases beyond the perturbative regime, indicating a transition to the strong-field regime. The selected harmonic spectra as a vertical line cut of 2D color map for zigzag (c) and armchair (d) orientations. The spectral evolution shows a broadening and emergence of additional features as the laser intensity increases, highlighting the critical role of their interference in non-trivial features in the non-perturbative regime.

**Supplementary Note**

**Sample fabrication**

A high-quality monolayer $WS_2$ sample was prepared by the gold tape exfoliation method [2]. A 150 nm thickness gold layer was deposited on a flat silicon substrate with a 90 nm thick oxide layer. A polyvinylpyrrolidone (PVP) solution (Sigma Aldrich, mw 40000, 10% wt in ethanol/acetonitrile wt 1/1) was spin-coated on the top of the Au film and cured at 150 °C for 5 minutes. This PVP layer served as a sacrificial layer to prevent contamination from tape residue. The prepared PVP/Au was picked up with thermal release tape (TRT), revealing an ultra-flat, clean, and fresh gold surface-referred to as the gold tape. The gold tape is pressed onto a freshly cleaved bulk $WS_2$ crystal (HQ graphene). As the tape is lifted off the surface, it carries the PVP/Au layer with a monolayer WS2 crystal attached to the Au surface. And then it is further transferred onto a silicon substrate with a 90 nm thick oxide layer. The thermal release tape is removed by heating at 135 °C. The PVP layer is removed by dissolving in deionized (DI) water for 2 hours. Finally, the sample on the substrate, covered by Au layer, was rinsed with acetone and cleaned by $O_2$ plasma for 4 minutes to remove any remaining polymer residues.

The van der Waals heterostructure of WS2 monolayer and hBN was prepared by the dry transfer technique [3]. Thickness of WS2 monolayer was first identified by the optical contrast of a microscope image, followed by the detailed spectroscopic characterization. Approximately 20 nm-thick hBN flakes were exfoliated onto a silicon substrate with 90 nm oxide layer. To fabricate the encapsulated $WS_2$ monolayer, we used the thermoplastic methacrylate copolymer (Elvacite 2552C, Lucite International) stamp to pick up the hBN flakes and $WS_2$ monolayer in sequence with accurate alignment based on an optical microscope. The Elvacite stamp with the heterostructure was then stamped onto a sapphire substrate. The polymer and samples were heated up at 70 °C for the pick-up and 200 °C for the stamp process, respectively. Finally, we dissolved the Elavcite in acetone for 3 minutes at 100 °C.

**Reflection contrast spectroscopy**

For reflection contrast spectra, a tungsten lamp was used as the broadband white light source. The incident light was focused onto the monolayer $WS_2$ in a home-built microscopy setup, and the reflected light was collected and analyzed using a spectrometer equipped with an electrically cooled Si CCD. To isolate the reflected light from the monolayer $WS_2$, a spatial filter (iris) was placed at the image plane in a 4f-system. The reflection contrast ($\Delta R/R_0$) spectrum was obtained by comparing the reflected light spectrum from the sample (R) with that from the substrate immediately adjacent to the sample ($R_0$), using the formula $\Delta R/R_0 = (R - R_0)/R_0$. Since reflection contrast is directly related to the imaginary part of $\varepsilon_2$ (optical absorption) of the material, it can be used as an approximation of the absorption spectrum.

**HHG measurements**

Mid-infrared pulses were generated from a femtosecond laser system (Light Conversion PHAROS) using an optical parametric amplifier (ORPHEUS) and a difference frequency generator (LYRA). The output served wavelength-tunable multi-cycle pulses with a repetition rate of 100 kHz. The spectral linewidth of the pulse was 15.4 meV in full width at half maximum (FWHM), and the pulse duration was estimated to be 120 fs, assuming a Fourier-transform-limited pulse. Laser intensity was precisely controlled by a pair of linear polarizers inserted into the beam path. Half-wave plates were also inserted into the beam path to control the polarization of the excitation laser. The mid-infrared pulses were then focused near the center of the monolayer $WS_2$ sample using ZnSe focusing objectives, producing a spot size of approximately 60–80 µm. The emitted HHG was collected by a 50× objective lens in a transmission geometry, and its polarization was analyzed by a half-wave plate mounted on a motorized stage and a fixed Glan-Taylor polarizer. The HHG spectra were recorded by an electron-multiplying charge-coupled device detector (ProEM, Princeton Instruments) and a grating spectrometer (SP-2300, Princeton Instruments).

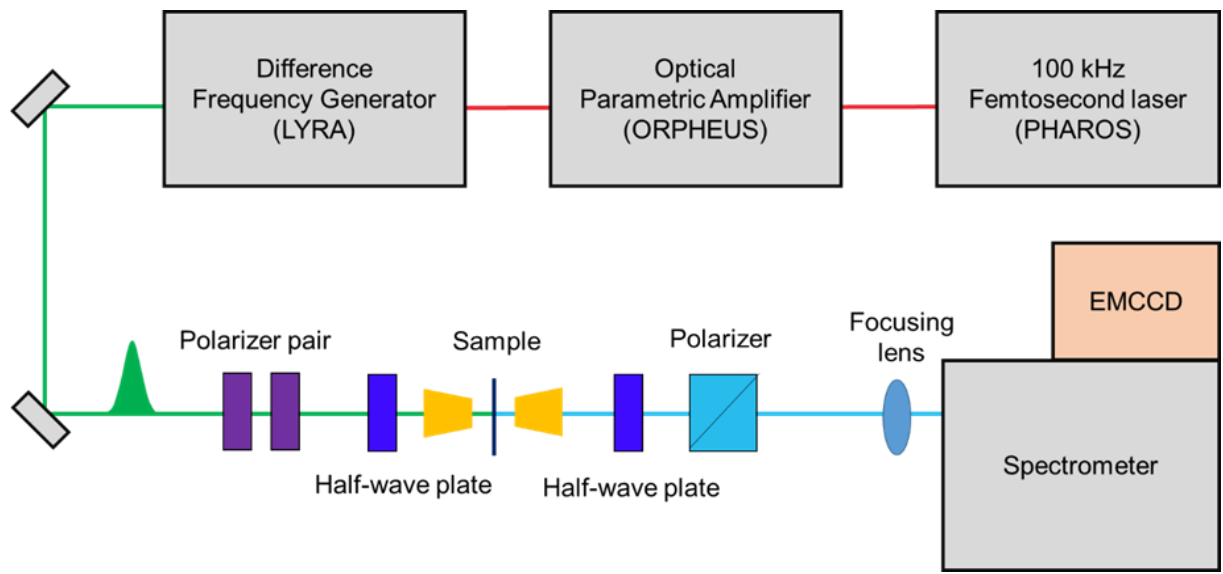

**Figure. S6| Schematic of the experimental setup for the HHG measurement.**

**Theoretical details**

**TDDFT simulations**

Time-dependent density functional theory (TDDFT) calculations are performed using the real-space grid-based code Octopus [4]. The Kohn-Sham (KS) equations were discretized on a 3D Cartesian grid within the primitive unit cell of monolayer $WS_2$ using the experimental lattice parameter a=3.186. A vacuum spacing of 30 Bohr is included above and below the monolayer to prevent spurious interactions.

Calculations are performed using the local density approximation (LDA), neglecting spin degrees of freedom and spin-orbit coupling. The frozen-core approximation is applied, with inner core states treated using norm-conserving pseudopotentials [5]. The KS equations are solved self-consistently with an energy convergence threshold of $10^{-7}$ Hartree, and the real-space grid spacing is converged to 0.35 Bohr. A Γ-centred *k*-grid with 36x36 *k*-points is used for BZ sampling.

In the TDDFT calculations, the time-dependent KS equations are solved within the adiabatic approximation in the velocity gauge and with an added complex absorbing potential at the edges of the z-axis of width 12 Bohr. In atomic units, the KS equations are given by

$$i\partial_t |\psi_{nk}(t)\rangle = \left[\frac{1}{2}\left(-i\nabla - \frac{\mathbf{A}(t)}{c}\right)^2 + v_{ks}(\mathbf{r},t)\right] |\psi_{nk}(t)\rangle \qquad (3)$$

where $|\psi_{nk}(t)\rangle$ are the KS single-particle Bloch states for band *n*, and *k*-point *k*, $\mathbf{A}(t)$ is the vector potential of the laser field within the dipole approximation, such that $-\partial_t \mathbf{A}(t) = c\mathbf{E}(t)$, *c* is the speed of light, and $v_{KS}(\mathbf{r},t)$ is the time-dependent KS potential given by

$$v_{ks}(\mathbf{r},t) = -\sum_I \frac{Z_I}{|\mathbf{R}_I - \mathbf{r}|} + \int d^3 r' \frac{n(\mathbf{r}',t)}{|\mathbf{r}-\mathbf{r}'|} + v_{xc}[n(\mathbf{r},t)] \qquad (4)$$

where $Z_I$ and $\mathbf{R}_I$ denote the charge and position of the *I*th nucleus, respectively. $v_{xc}$ represents the exchange-correlation (XC) potential, which is a functional of the time-dependent electron density n(**r**,t). In practice, the Coulomb interaction with the nuclei is replaced by a non-local pseudopotential, which also accounts for core electron contributions.

In all calculations we assumed that the nuclei are static. In calculations invoking the independent particle approximation (IPA) we froze the temporal-dependence of the KS potential to its ground state form, i.e. $v_{ks}(\mathbf{r}, t) = v_{ks}(\mathbf{r}, t = 0)$, which decouples the TDDFT equations of motion for different Bloch states, and leads to independent electrons that due not dynamically interact (i.e. electron-electron interactions are not dynamically evolving). From these simulations we calculated the total current in a method similar to that in refs. [6, 7], from which we found the resulting HHG spectra as described in the main text. The time-dependent equations were solved with a time step of $\Delta t = 0.2\ a.u.$ We also projected the electronic occupations on the ground state KS orbitals to evaluate the level of contributions of different valence and conduction bands to the simulation with an approach as in refs. [6, 7]. We conducted simulations at similar laser conditions to those in the main text.

Our main results are shown in fig. S7, clearly indicating that the IPA is fully applicable in our conditions (at least up to the 15$^{th}$ harmonic order), and that the dominant electron occupation is induced in the first valence and first conduction bands (with the percentage of active carriers in those bands being ~72.5% out of the full excitation). These results motivate and validate the application of the SBE approach, in which higher (and lower) lying bands, as well as electron-electron interactions, are neglected. The HHG spectra from the TDDFT simulations themselves were shown to not reproduce the main experimental features due to high level of noise, which in these conditions arises due to lack of proper dephasing in the velocity gauge in conditions where the simulation time is very long.

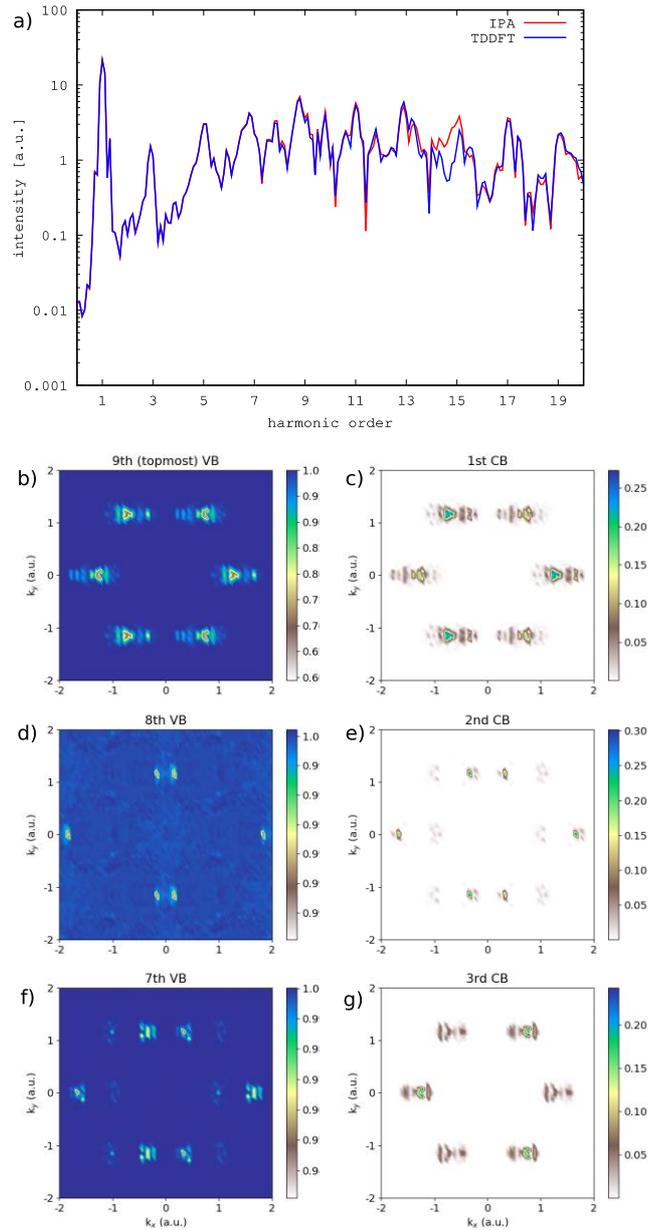

**Figure S7| TDDFT simulations of HHG from WS$_2$.** (a) HHG spectra with/without the IPA, in typical experimental conditions, clearly showing the IPA is valid. (b-g) projected electron/hole occupation at the end of the simulation onto WS$_2$ conduction/valence bands (up to 3rd conduction band, and down to third valence band from the topmost (numbered 9$^{th}$)). The dominant occupation is induced in the 1st valence/conduction bands, in which the sum of free carriers amounts to ~72.5% of the total free carriers. These two results motivate and validate the application of SBE simulations with proper dephasing. Simulations here employed a 10-cycle laser field at a super-sine shape. The employed envelope function is taken as a super-sine form [8] at intensity of 200 GW/cm$^2$.

We formulate a TB model for WS$_2$ based on an AB honeycomb lattice structure (neglecting out-of-plane components). The TB Hamiltonian in the AB sublattice basis reads:

$$H_0 = \frac{\Delta}{2}\sigma_z + \begin{pmatrix} \alpha(\mathbf{k}) & \beta(\mathbf{k}) \\ \beta^*(\mathbf{k}) & \alpha(\mathbf{k}) \end{pmatrix} \quad (5)$$

where $\sigma_z$ is the z Pauli Matrix, $\Delta$ is the optical gap at K/K' which is taken at the experimental value (resonant with the 7-photon transition discussed in the main text), and $\alpha$ and $\beta$ are structure factors determined by the hopping parameters (up to 14$^{th}$ order here) between various sublattice sites:

$$\begin{aligned}\alpha(\mathbf{k}) &= \begin{pmatrix} t_2 f_2(\mathbf{k}) + t_5 f_5(\mathbf{k}) + t_6 f_6(\mathbf{k}) \\ + t_{10} f_{10}(\mathbf{k}) + t_{11} f_{11}(\mathbf{k}) + t_{14} f_{14}(\mathbf{k}) \end{pmatrix} \\ \beta(\mathbf{k}) &= \begin{pmatrix} t_1 f_1(\mathbf{k}) + t_3 f_3(\mathbf{k}) + t_4 f_4(\mathbf{k}) + t_7 f_7(\mathbf{k}) \\ + t_8 f_8(\mathbf{k}) + t_9 f_9(\mathbf{k}) + t_{12} f_{12}(\mathbf{k}) + t_{13} f_{13}(\mathbf{k}) \end{pmatrix}\end{aligned} \quad (6)$$

, with $f_n(\mathbf{k}) = \sum_j \exp(i\mathbf{k}\cdot \mathbf{h_{nj}})$ the structure factors, where $\mathbf{h_{nj}}$ is the hopping vector between nearest-neighbor (NN) sites of order $n$. This Hamiltonian is diagonalized to obtain the band energies $\varepsilon_\pm = \alpha(\mathbf{k}) \pm \sqrt{|\beta(\mathbf{k})|^2 + \frac{\Delta^2}{4}}$, as well as eigenfunctions. The band energies are fitted throughout the full BZ with a strategy similar to that in ref. [7], and with a double weight near the K/K' valleys to obtain an optimal agreement of the electronic structure in that region (including Berry curvature). This provides the optimal set of hopping parameters, $t_j$, employed in structure factors. Note that this model neglects spin degrees of freedom, as well as out-of-plane structure and various hybridized orbital character of TMD bands. Nonetheless, it can capture the underlying physics of the valleys, and especially the correct band structure throughout the BZ including band inverted regions near Γ and M (see Fig. Sxx comparing DFT and TB bands). Especially, we note that this is enabled by using such a high order TB model (while typical applications use 2$^{nd}$ or 3$^{rd}$ order neighbor hopping), which replaces the need to employ multiple bands by added long range information. In Fig. 4 in the main text we also employ a reduced TB model with only up to 3$^{rd}$ order terms, and where the bands are optimally fitted only in the region of the K/K' valleys, yielding much larger optical gaps away from the valleys.

From the band energies and eigenstates, we analytically obtain the transition matric dipole elements (TDME) employed in the main text by: $\mathbf{d}_{ij}(\mathbf{k}) = i\langle\psi_i(\mathbf{k})|\frac{\partial}{\partial \mathbf{k}}|\psi_j(\mathbf{k})\rangle$, and the momentum matrix elements: $\mathbf{p}_{ij}(\mathbf{k}) = \langle\psi_i(\mathbf{k})|\frac{\partial H_0(\mathbf{k})}{\partial \mathbf{k}}|\psi_j(\mathbf{k})\rangle$. These are employed in the SBE simulations discussed in the main text, which are solved using a 4$^{th}$-order Runge Kutte method with a time step of 0.02 a.u., and by assuming initial conditions where electrons fully occupy the VB. In these simulations we employ a laser field in accordance with that employed in the experiments, only with a different duration and envelope function. The employed envelope function is taken as a super-sine form [8], with a full-width-half-max of 60 femtoseconds. For the simulations that require much better spectral resolution discussed in the main text (resolving the peak splitting phenomena in the 7$^{th}$ harmonic), much longer duration driving pulses are employed in order to obtain sufficient spectral resolution (FWHM 375 femtoseconds). Note that such long timescale dynamics are necessary to be able to resolve peal splitting on an energy scale of ~0.01 eV, while in experiments this resolution is readily obtained at the spectrometer. Integrated harmonic yields are obtained by integrating the HHG emission around the harmonic peak, and k-resolved harmonic yields are obtained by integration around the harmonic peak of HHG emission contributed from the current of a particular k-point (i.e. avoiding the *k*-summation in eq. (2) in the main text).

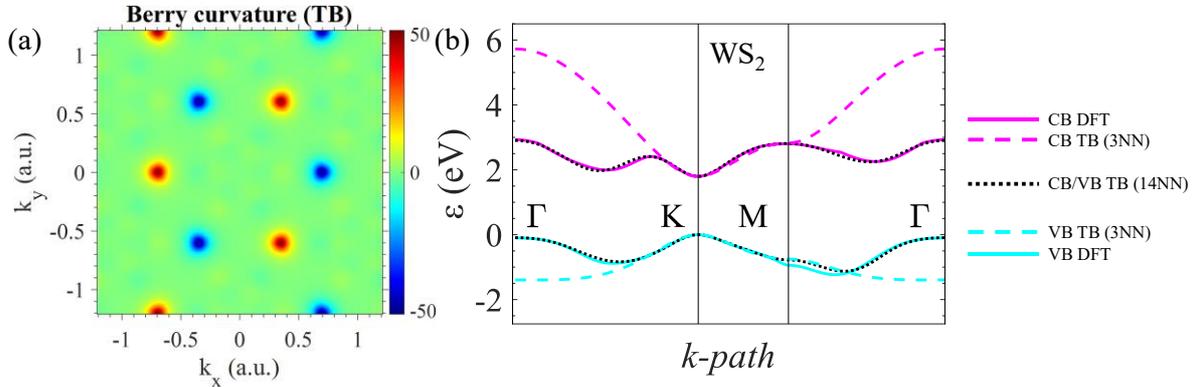

**Figure S8| TB model electronic structure.** (a) Optimally fitted (14NN) TB model obtained Berry curvature in the BZ, showing correct valley region Berry curvature. (b) TB bands compared to DFT bands along high symmetry lines. The extended TB model (14NN) fit very well throughout the BZ, yielding good optical gaps. A 3$^{rd}$-order NN TB model only obtains good bands near the K/K' valleys, with much higher optical gaps towards Γ.

**References**

[1]     Vampa, G. *et al.* Theoretical Analysis of High-Harmonic Generation in Solids. *Phys. Rev. Lett.* **113**, 073901 (2014).

[2]     Liu, F. *et al.* Disassembling 2D van der Waals crystals into macroscopic monolayers and reassembling into artificial lattices. *Science* **367**, 903–906 (2020).

[3]     Wang, L. *et al.* One-Dimensional Electrical Contact to a Two-Dimensional Material. *Science* **342**, 614–617 (2013).

[4]     Tancogne-Dejean, N. et al. Octopus, a computational framework for exploring light-driven phenomena and quantum dynamics in extended and finite systems. *J. Chem. Phys.* **152**, 124119 (2020).

[5]     Hartwigsen, C., Goedecker, S. & Hutter, J. Relativistic separable dual-space Gaussian pseudopotentials from H to Rn. *Phys. Rev. B* **58**, 3641 (1998).

[6]     Neufeld, O. *et al*, A. Light-driven extremely nonlinear bulk photogalvanic currents. *Phys. Rev. Lett*. **127**, 126601 (2021).

[7]     Galler, A., Rubio, A. & Neufeld, O. Mapping Light-Dressed Floquet Bands by Highly Nonlinear Optical Excitations and Valley Polarization. *J. Phys. Chem. Lett.* **14**, 11298–11304 (2023).

[8]     Neufeld, O. & Cohen, O. Background-free measurement of ring currents by symmetry-breaking high-harmonic spectroscopy. *Phys. Rev. Lett.* **123**, 103202 (2019).